\documentclass[12pt]{iopart}
\pdfoutput=1
\usepackage{amssymb}
\usepackage{amsfonts}
\usepackage{setstack}
\usepackage{amsthm}
\usepackage{epsfig}
\usepackage{graphicx}
\usepackage{color}
\usepackage[latin1]{inputenc}

\newcommand{\lrangle}[1]{\langle{#1}\rangle}

\newcommand{\pkk}{\psi_{kk'}}
\newcommand{\okk}{\omega_{kk'}}
\newcommand{\fkk}{\phi_{kk'}}
\newcommand{\bfkk}{\bar{\phi}_{kk'}}

\begin{document}
\title[Heterogeneous pair-approximation for the contact process on complex
networks]{Heterogeneous pair-approximation for the contact process on complex
networks}
\author{Ang\'elica S. Mata$^1$\footnote{On leave at Departament de F\'{\i}sica i Enginyeria Nuclear,
  Universitat Polit\`ecnica de Catalunya, Barcelona, Spain},
Ronan S. Ferreira$^2$ and
Silvio C. Ferreira$^1$}
\ead{angelica.mata@ufv.br,ronan.ferreira@ua.pt,silviojr@ufv.br}
\address{$^1$Departamento de F\'{\i}sica, Universidade Federal de Vi\c{c}osa, 
36570-000, Vi\c{c}osa, MG, Brazil}
\address{$^2$
Department of Physics \& I3N, University of Aveiro, 3810-193 Aveiro, Portugal}
\date{\today}

\begin{abstract}
Recent works have shown that  the contact process running on the top of highly
heterogeneous random networks is described by the heterogeneous mean-field
theory. However, some important aspects as the transition point and strong
corrections to the finite-size scaling observed in simulations are not
quantitatively reproduced in this theory. We develop a heterogeneous 
pair-approximation, the simplest mean-field approach that takes into account
dynamical correlations, for the contact process. The transition points obtained
in this theory are in very good agreement with simulations. The proximity with a
simple homogeneous pair-approximation is elicited showing that the transition
point in successive homogeneous cluster approximations moves away from the
simulation results. We show that the critical exponents of the heterogeneous
pair-approximation in the infinite-size limit are the same as those of the
one-vertex theory. However, excellent matches with simulations, for a wide range
of network sizes, are obtained when sub-leading finite-size corrections given by
the new theory are explicitly taken into account. The present approach  can be
suited to dynamical processes on networks in general providing a profitable
strategy to analytically assess fine-tuning theoretical corrections.

\end{abstract}


\section{Introduction}

The accurate theoretical understanding of dynamical systems in the form of
reaction-diffusion processes running on the top of complex networks rates among
the hottest issues in complex network theory~\cite{Goltsev12, VanMieghem12,
Castellano10,odor2013spectral, Lee2013, boguna2013nature,
mata2013pair,Pugliese09, Gomez10, Gomez:nopert, JuhaszCP,Castellano:2008,
Hong2007}. Much effort has been devoted to the criticality of the ensuing
absorbing state phase transition observed in the contact process
(CP)~\cite{JuhaszCP,Castellano:2008, Hong2007, Boguna09, cpannealed} and in the
susceptible-infected-susceptible (SIS)~
\cite{Goltsev12,Castellano10,odor2013spectral, Lee2013,
boguna2013nature, mata2013pair,Gomez:nopert} models, mainly based on perturbative approaches
around the transition point~\cite{Goltsev12, VanMieghem12,
Castellano10,odor2013spectral,mata2013pair,Gomez10,Castellano:2008}, even though
non-perturbative analyses have recently been performed~\cite{Gomez:nopert}. 

The heterogeneous mean-field (HMF) approach for dynamical processes on complex
networks has become widespread in the last years. This theory, formerly
conceived to investigate the SIS dynamics on complex networks~\cite{Romu},
assumes that the number of connections of a vertex (the vertex degree) is the
quantity relevant to determine its  state, neglects all dynamical
correlations as well as the actual structure of the network. On the other hand,
the quenched mean-field (QMF) theory~\cite{Castellano10,Wang03} still neglects
dynamical correlations but the actual quenched structure of the network is
explicitly taken into account by means of the adjacency matrix $A_{ij}$ that
contains the complete information of the connection among
vertices~\cite{Newman10}. More recently, semi-analytic methods including
dynamical fluctuations~\cite{Lee2013,boguna2013nature} and heterogeneous
pair-approximations~\cite{Pugliese09,Cator12,Gleeson2013,mata2013pair} have
appeared as more accurate alternatives to HMF theory.

The CP~\cite{harris74} is the simplest reaction-diffusion process exhibiting a
transition between an active and a frozen (absorbing) phase~\cite{marro1999npt}.
The CP dynamics investigated in the present work is defined as
follows~\cite{marro1999npt}: A vertex $i$ of an arbitrary unweighted graph can
be occupied ($\sigma_i=1$) or empty ($\sigma_i=0$). At a rate $\lambda$,
an occupied vertex tries to create an offspring in a randomly chosen
nearest-neighbor, what happens only if it is empty. An occupied vertex
spontaneously disappears at rate $1$ (this rate fixes the time unit). Notice
that in the SIS dynamics an occupied vertex creates (`infects' in the epidemiological
jargon) an offspring in each empty nearest neighbor at rate $\lambda$. 
Even being equivalent for strictly
homogeneous graphs ($k_i\equiv k~\forall~i$), these models are very different for
heterogeneous substrates (see discussion in Ref.~\cite{sander_phase_2013}).
However, in both models the creation of particles is a catalytic process
occurring exclusively in pairs of empty-occupied vertices, implying that the
state devoid from particles is a fixed point of the dynamics and is called 
absorbing state. 

After an intense discussion~\cite{Hong2007, Castellano:2008,Castellano:2006,
comment1, comment2}, the HMF theory showed up as the best available approach to
describe scaling exponents associated to the phase transition of the CP on
networks~\cite{cpquenched}. However, some important questions remained
unanswered. The transition point $\lambda_c=1$ predicted
by the HMF theory~\cite{Castellano:2006} does not reflect the dependence on the degree distribution
observed in simulations~\cite{JuhaszCP,Castellano:2006,cpquenched}. Most
intriguingly, it was observed a good accordance between simulations and a
heuristic modification of the strictly homogeneous pair-approximation (HPA) (see
Ref.~\cite{marro1999npt} for a review) where the fixed vertex degree is replaced
 by the average degree of the network~\cite{JuhaszCP,cpquenched}:
\begin{equation}
 \lambda_c=\frac{\lrangle{k}}{\lrangle{k}-1}.
\label{eq:lbc_hom}
\end{equation}
Finally, sub-leading corrections to the finite-size scaling, undetected by the
one-vertex HMF theory, are quantitatively relevant for the analysis of highly
heterogeneous networks, for which deviations from the theoretical
finite-size  scaling exponents were reported~\cite{cpquenched}.

Dynamical correlations represent an important
factor to ascertain the accuracy of the analytical results. The simplest way to
explicitly consider dynamical correlations is by means of a
pair-approximation~\cite{marro1999npt}. In this paper, we present a pair
HMF approximation for the CP on heterogeneous networks.
We show that this theory yields great improvements in relation to the
one-vertex counterpart but, however,  may be farther from the simulation
thresholds than Eq.~(\ref{eq:lbc_hom}). 
This apparent contradiction is solved showing that the higher-order
homogeneous cluster approximations overestimate the actual transition point
implying that the proximity is only a coincidence. We also show that pair HMF
theory yields  the same critical exponents as the one-vertex HMF theory, but
with different corrections to the scaling. These corrections allow an almost
perfect match with simulations constituting a great improvement in relation to the
one-vertex mean-field theories~\cite{cpannealed,Boguna09}.

The paper is organized  as follows: Pair HMF theory is proposed and
the transcendental equation that gives the transition points is derived in
section \ref{sec:phmf}. The numerical analyses of the thresholds and the
comparisons with quasi-stationary simulations are presented in
section~\ref{sec:numerics}. The critical exponents are analytically determined
and compared with simulations in
section~\ref{sec:critical}. Our concluding remarks are drawn in
section~\ref{sec:conclu}.

\section{Pair HMF theory}
\label{sec:phmf}

In this section, we develop the pair HMF theory where the evolution  of the
system is given by the average behavior of vertices  with the same
degree. So, let us introduce the notation based on Ref.~\cite{mata2013pair}: $[A_k]$ is the probability that a
vertex of degree $k$ is in the state $A$; $[A_kB_{k'}]$ is the probability that
a vertex of degree $k$ in state $A$ is connected to a vertex of degree $k'$  in
state $B$; $[A_k B_{k'} C_{k''}]$ is the generalization to three vertices such
that the pairs $[A_k B_{k'}]$ and $[B_{k'} C_{k''}]$ are connected through a
node of degree $k'$ and so forth. An occupied state is represented by $1$ and an
empty one by $0$. The pair-approximation carried out hereafter uses the
following notation: $[1_k] = \rho_k$, $[0_k] = 1-\rho_k$, $[0_k1_{k'}]=\fkk$,
$[1_k0_{k'}]=\bfkk$, $[1_k1_{k'}]=\pkk$ and $[0_k0_{k'}]=\okk$. Obviously, we
have that $\psi_{kk'}=\psi_{k'k}$, $\omega_{kk'}=\omega_{k'k}$, and
$\phi_{kk'}=\bar{\phi}_{k'k}$. Independently of the dynamical rules, the
following closure relations can be derived from simple probabilistic reasonings:
\begin{eqnarray}
\psi_{kk'}+\phi_{kk'} & = &\rho_{k'}\nonumber \\ 
\psi_{kk'}+\bar{\phi}_{kk'} & = &\rho_k \nonumber\\ 
\omega_{kk'}+\phi_{kk'} & = &1-\rho_k \nonumber \\ 
\omega_{kk'}+\bar{\phi}_{kk'}& = &1-\rho_{k'}  
\label{eq:norms}.
\end{eqnarray} 
The master equation for the probability that a vertex with degree $k$ is
occupied takes the form
\begin{equation}
\frac{d\rho_k}{dt} = -\rho_k+\lambda k\sum_{k'}\frac{\fkk}{k'}P(k'|k),
 \label{eq:rhok1}
\end{equation}
where the conditional probability $P(k'|k)$, which gives the probability that a
vertex of degree $k$ is connected to a vertex of degree $k'$, weighs the
connectivity between compartments of degrees $k$ and $k'$.
The first term of Eq.~(\ref{eq:rhok1}) represents the spontaneous annihilation 
and the second term reckons the creation in a vertex of degree $k$ due to its
nearest neighbors. The dynamical equation for $\fkk$  is
\begin{eqnarray}
\frac{d\fkk}{dt} & = & -\fkk-\lambda\frac{\fkk}{k'}+\pkk+ 
\lambda(k'-1)\sum_{k''}\frac{[0_k0_{k'}1_{k''}]P(k''|k')}{k''}\nonumber \\
 & &-\lambda(k-1)\sum_{k''}\frac{[1_{k''}0_{k}1_{k'}]P(k''|k)}{k''}.
 \label{eq:phikk1}
\end{eqnarray}
The first term
represents the annihilation in the vertex of degree $k'$, the second one includes
the creation in the vertex of degree $k$ due to the connection with the neighbor 
of degree $k'$ and the third
one is due to the annihilation of the vertex with degree $k$. These terms
represent the reactions inside pairs with degrees $k$ and $k'$, that
create or destroy a configuration $[0_k,1_{k'}]$. The fourth and fifth terms
represent changes due to creation in vertices with degree $k'$ and $k$,
respectively, due to all their neighbors except the link between the vertices of
the pair itself, which is
explicitly included in the second term.

The one-vertex mean-field equation proposed in Ref.~\cite{Castellano:2006} is
obtained factoring the joint probability $\phi_{kk'}\approx (1-\rho_k)\rho_{k'}$
in Eq.~(\ref{eq:rhok1}). Details of one-vertex solution can be found
elsewhere~\cite{Boguna09}. Finally, the factor $k'-1$ preceding the first
summation in Eq.~(\ref{eq:phikk1})  is due to the $k'$ neighbors of middle
vertex except the link of the pair $[0_k0_{k'}]$ (similarly for $k-1$ preceding
the second summation).

We now approximate the triplets in Eq.~(\ref{eq:phikk1}) with a standard
pair-approximation \cite{Dickman88,Avraham92,henkel2008non}
\begin{equation}
 [A_k,B_{k'},C_{k''}] \approx \frac{[A_k,B_{k'}][B_{k'},C_{k''}]}{[B_{k'}]},
\end{equation}
to find
\begin{eqnarray}
\frac{d\fkk}{dt} & = & -\fkk-\lambda\frac{\fkk}{k'}+\pkk + 
\frac{\lambda(k'-1)\okk}{1-\rho_{k'}}\sum_{k''}\frac{\phi_{k'k''}
P(k''|k')}{k''}\nonumber \\
& & - 
\frac{\lambda(k-1)\fkk}{1-\rho_k}\sum_{k''}\frac{\phi_{kk''}P(k''|k)}{k''}.
 \label{eq:phikk2}
\end{eqnarray}
Substituting Eqs.~(\ref{eq:norms}) in (\ref{eq:phikk2}) and performing a linear 
stability analysis around the fixed point $\rho_k \approx 0$ and $\fkk \approx 0$,
one finds
\begin{equation}
 \frac{d\fkk}{dt} = -\left(2+\frac{\lambda}{k'}\right)\fkk+\rho_{k'}+
 \lambda(k'-1)\sum_{k''}\frac{\phi_{k'k''}P(k''|k')}{k''}.
 \label{eq:phikklin1}
\end{equation}
The next step is to perform a quasi-static approximation for
$t\rightarrow\infty$, in which $d\rho_k/dt \approx 0$ and $d\fkk/dt \approx 0$, 
to find
\begin{equation}
 \fkk = \frac{2k'-1}{2k'+\lambda}\rho_{k'}.
 \label{eq:fkkO1}
\end{equation}
Finally, we plug Eq.~(\ref{eq:fkkO1}) in Eq.~(\ref{eq:rhok1}) to produce
\begin{equation}
 \frac{d\rho_k}{d t} =\sum_{k'}L_{kk'}\rho_{k'},
\end{equation}
 where the Jacobian $L_{kk'}$ is given by
\begin{equation}
L_{kk'} = -\delta_{kk'}+  \frac{\lambda k(2k'-1)P(k'|k)}{(2k'+\lambda)k' }
 = -\delta_{kk'}+C_{kk'},
\label{eq:Jaco2} 
\end{equation}
with $\delta_{kk'}$ being the Kronecker delta symbol.

The absorbing state is unstable when the largest eigenvalue of $L_{kk'}$ is
positive. Therefore, the critical point is obtained when the largest eigenvalue
of the Jacobian matrix is null. Let us focus only on uncorrelated networks where
$P(k'|k)=k'P(k')/\lrangle{k}$~\cite{mariancutofss}. It is easy to check that
$u_k=k$ is an eigenvector of $C_{kk'}$ with eigenvalue
\begin{equation}
 \Lambda =
\frac{\lambda}{\lrangle{k}}\sum_{k'}\frac{(2k'-1)P(k')k'}{(2k'+\lambda)}.
 \label{eq:Lambda}
\end{equation}
Since $C_{kk'}>0$ is irreducible (all compartments have non-null chance of being
connected) and $u_k>0$, the Perron-Frobenius theorem~\cite{Newman10} warranties that $\Lambda$
is the largest eigenvalue of $C_{kk'}$. The transition point is, therefore,
given by $-1+\Lambda=0$ that results the transcendent equation
\begin{equation}
 \frac{\lambda_c}{\lrangle{k}}\sum_{k'}\frac{(2k'-1)k'P(k')}{(2k'+\lambda_c)}=1,
 \label{eq:lbc}
\end{equation}
which can be  numerically solved for any
kind of network (section~\ref{sec:numerics}).

To check the consistency of the theory, we consider the random regular networks
(RRNs) that are strictly homogeneous networks with vertex degree distribution
$P(k)=\delta_{k,m}$ and connections done at random avoiding self and multiple
edges~\cite{Ferreira12}. Upon substitution of  $P(k)$ in Eq.~(\ref{eq:lbc}), one
easily shows that the transition point is
\begin{equation}
 \lambda_c=\frac{m}{m-1},
\label{eq:lbc_RR}
\end{equation}
that is the same of the simple homogeneous pair-approximation.
Simulations of CP on RRNs with $m=6$ yield a critical point 
$\lambda_c=1.2155(1)$~\cite{ronan}, slightly above the pair-approximation 
prediction $\lambda_c=1.2$.

\section{Threshold for arbitrary  random networks}
\label{sec:numerics}

In this section, we compare the thresholds given by Eq.~(\ref{eq:lbc}) with
simulations of the CP dynamics  on random networks generated by the
uncorrelated configuration model (UCM)~\cite{Catanzaro05}. Power law degree
distributions $P(k)\sim k^{-\gamma}$, where $\gamma$ is the degree exponent, with
minimum degree $k_0$ and structural cutoff $k_c=N^{1/2}$, the latter rendering
networks without degree correlations~\cite{mariancutofss}, were used. This
choice is suitable for comparison with the pair HMF theory where such a
simplification was adopted. We investigated networks with either $k_0=3$ or $6$.
The latter is to compare with the results of Ref.~\cite{cpquenched} and to
remark the improvement of the present theory. Networks of sizes up to $N=10^7$
and degree exponents $\gamma=2.3,~2.5,~2.7,~3.0$ and 3.5 were analyzed.

\begin{figure}[ht]
\centering
 \includegraphics[width=7cm]{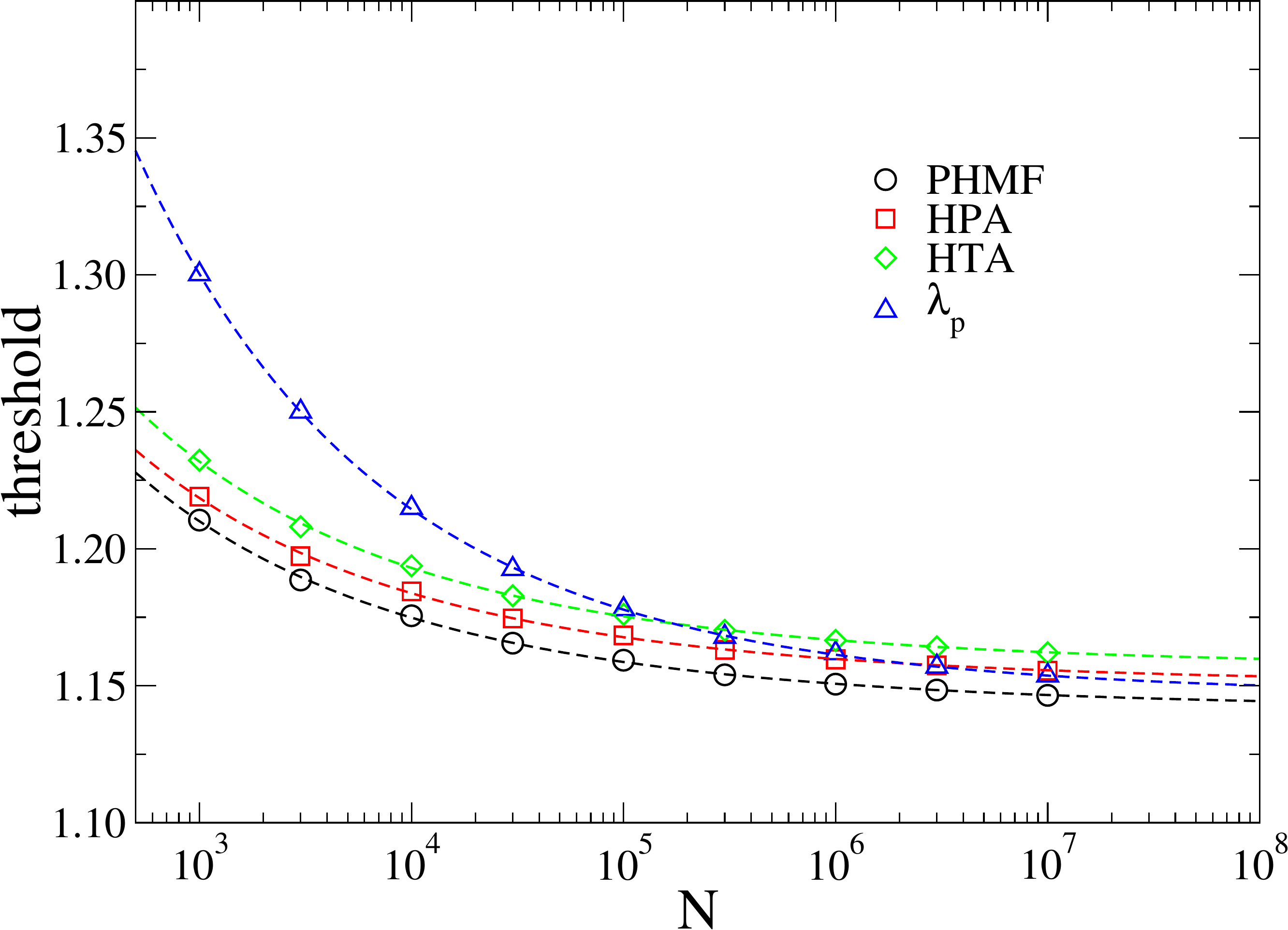} 
  \includegraphics[width=7.0cm]{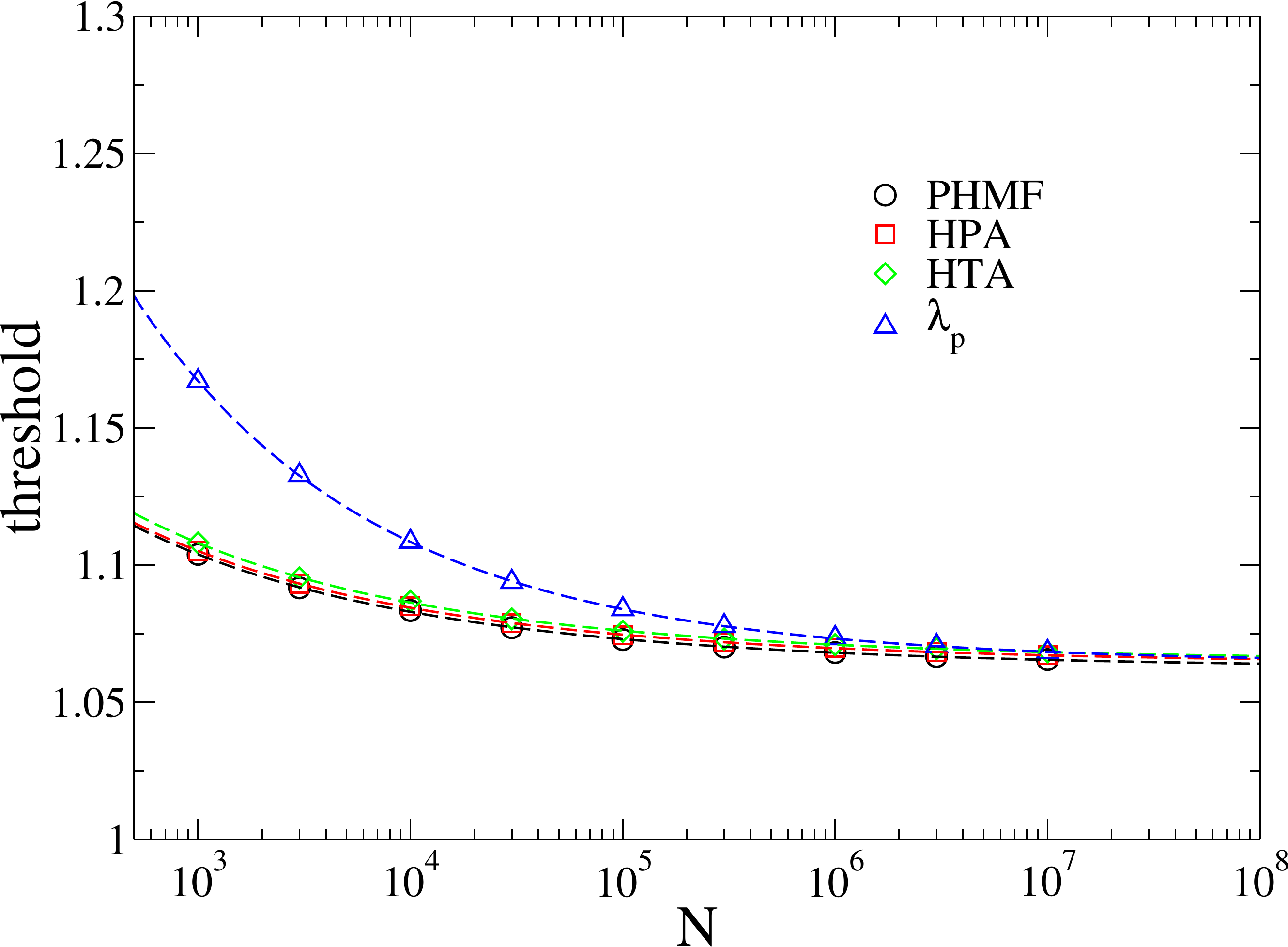}\\ ~ \\
 \includegraphics[width=7cm]{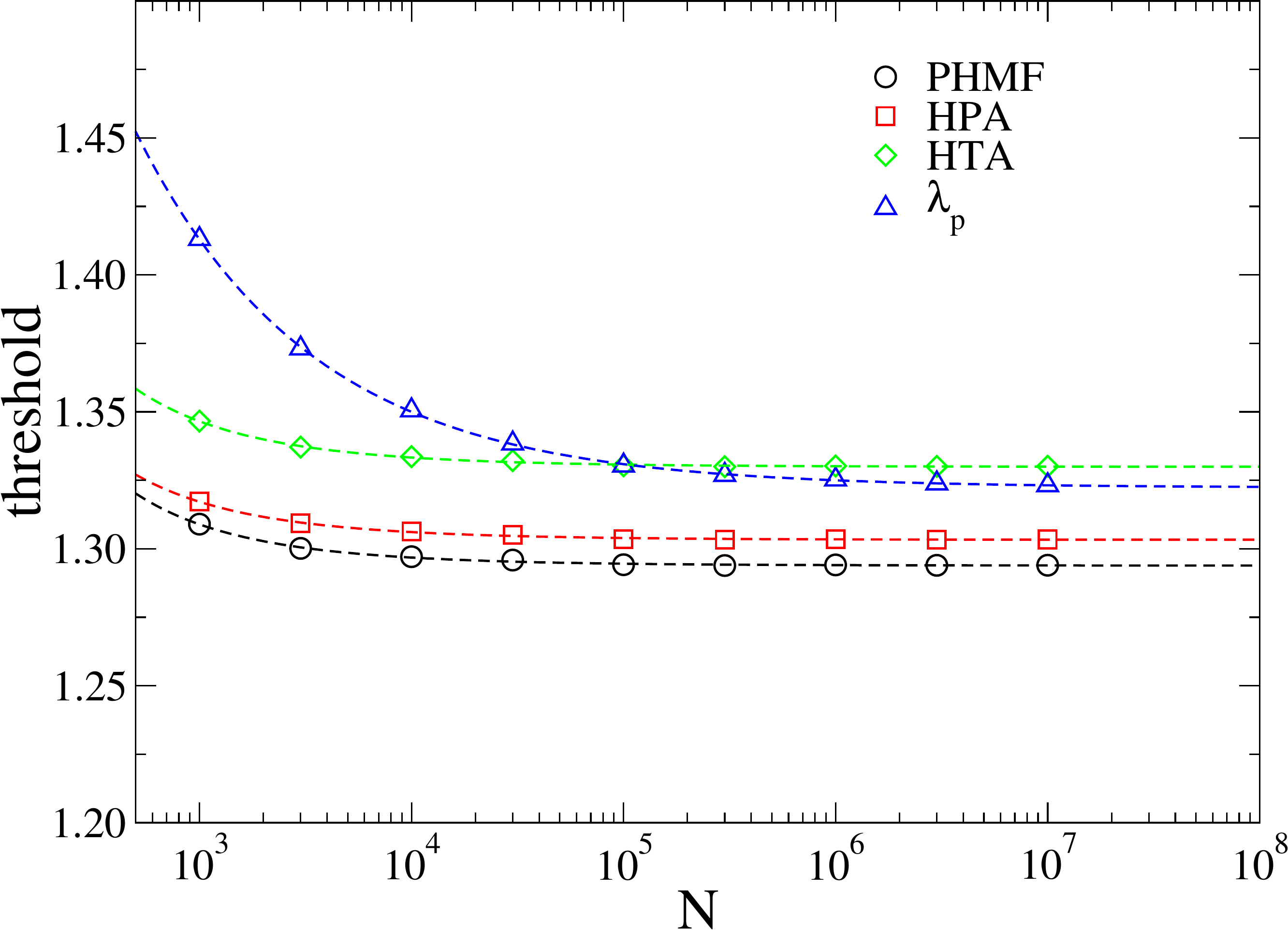} 
  \includegraphics[width=7.0cm]{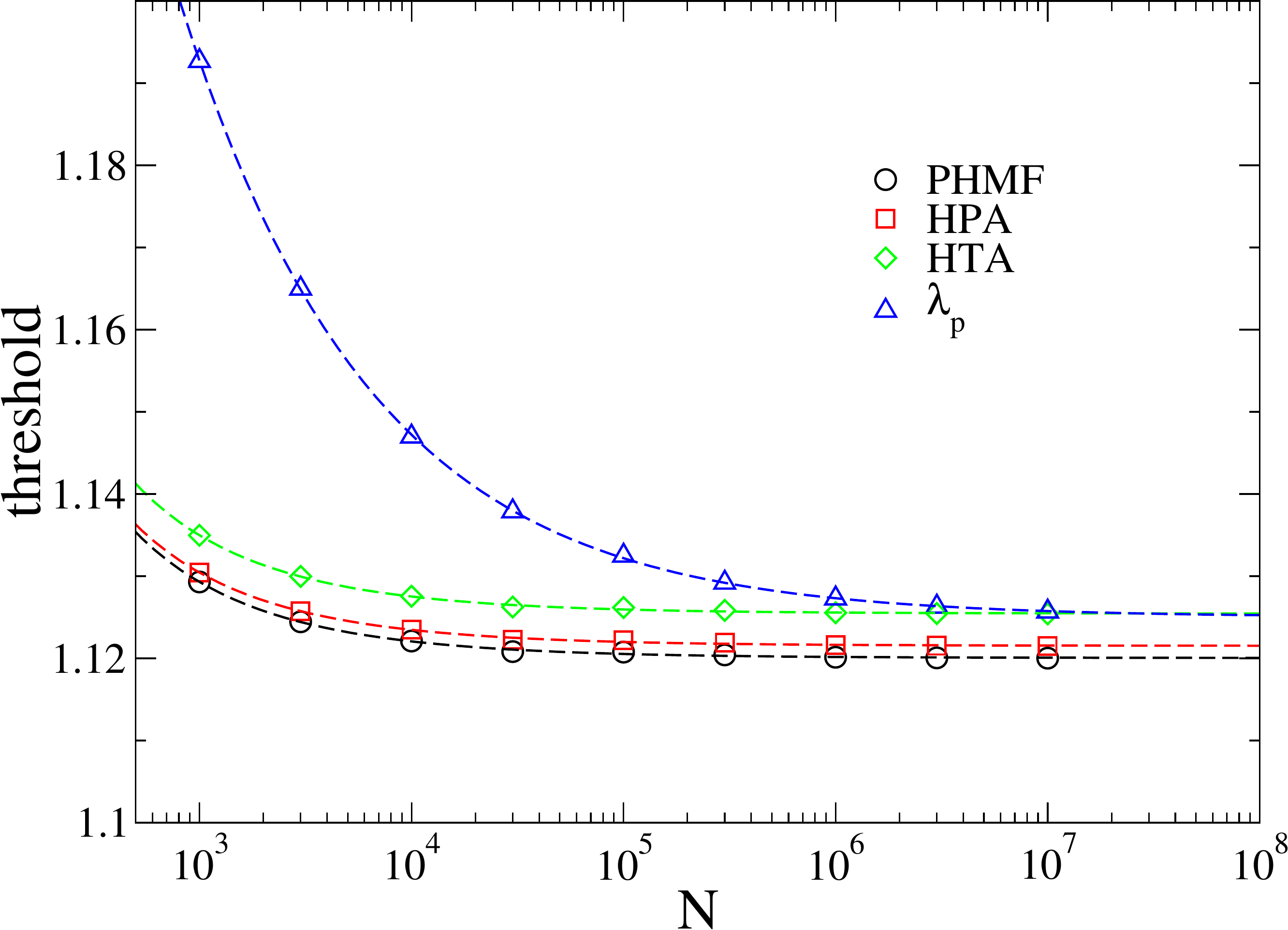}
 \caption{Thresholds against network size for the CP on UCM networks with degree
exponents $\gamma=2.50$ (top) and $\gamma=3.50$ (bottom), $k_0=3$ (left) and
$k_0=6$ (right), obtained in mean-field theories and QS simulations (position of
the susceptibility peak $\lambda_p$). Dashed lines are non-linear regressions,
Eq.(\ref{eq:nonlin}), used to extrapolate the infinite-size limit of the
thresholds. Acronyms: PHMF (pair heterogeneous mean-field, HPA (homogeneous 
pair approximation), HTP (homogeneous triplet approximation).}
 \label{fig:lbc}
\end{figure}

The thresholds for heterogeneous pair-approximations were determined for each
network realization and averages done over 10 independent networks.
Sample-to-sample fluctuations of the threshold positions become very small for
large networks. The  thresholds against network size for two degree exponents
are shown in Fig.~\ref{fig:lbc}. The results are compared with the heuristic
formula inspired in the HPA theory given by
Eq.~(\ref{eq:lbc_hom}). 
We performed simulations of CP dynamics on the same network samples used to
evaluate the  mean-field theories. The standard simulation scheme was
used~\cite{marro1999npt}: An occupied vertex $j$ is randomly chosen. With
probability $p=1/(1+\lambda)$ the selected vertex becomes vacant. With
complementary probability $1-p$ one of the $k_j$ nearest-neighbors  of $j$ is
randomly chosen and, if empty, is occupied. The time is incremented by $\Delta
t=1/[(1+\lambda)n(t)]$, where $n(t)$ is the number of particles at time $t$. To
overcome the difficulties intrinsic to the simulations of systems with absorbing
states~\cite{marro1999npt}, we used the quasi-stationary (QS) simulation
method~\cite{Mancebo2005}, in which every time the system tries to visit an
absorbing state it jumps  to an active configuration
previously visited during the simulation (a new initial condition). Details of the method with
applications to dynamical processes on networks  can be found
elsewhere~\cite{cpannealed,ronan}.

The QS probability $\bar{P}(n)$, defined as the probability that the system has
$n$ occupied vertices in the QS regime, was computed after a relaxation
$t_r=10^6$ during an averaging time $t_a=10^7$.
The transition point for finite networks was determined using 
the modified susceptibility~\cite{Ferreira12}
\begin{equation}
 \chi\equiv\frac{\lrangle{n^2}-\lrangle{n}^2}{\lrangle{n}}=
\frac{N(\lrangle{\rho^2}-\lrangle{\rho}^2)}{\lrangle{\rho}},
\label{eq:chi}
\end{equation}
which is expected to have a diverging peak that converges to the transition
point when the network size increases. 

The choice of the alternative definition, Eq.~(\ref{eq:chi}), instead of the
standard susceptibility $\tilde{\chi}=N(\lrangle{\rho^2}-\lrangle{\rho^2})$~\cite{henkel2008non} is
due to the peculiarities of dynamical processes on  complex networks.
For example, the CP on annealed networks\footnote{In annealed networks, the
vertex degrees are fixed while the edges are completely rewired between
successive dynamics steps implying that dynamical correlations are
absent and HMF theory becomes an exact prescription~\cite{Boguna09}.}, for
which the QS probability distribution at the transition point has the 
analytically known form~\cite{cpannealed}
\begin{equation}
 \bar{P}(n)=\frac{1}{\sqrt{\Omega}}f\left(\frac{N}{\sqrt{\Omega}}\right),
\label{eq:pofn}
\end{equation}
where $\Omega=N/g$, $g=\lrangle{k^2}/\lrangle{k}^2$ and $f(x)$ is a scaling
function independent of the degree distribution. It is easy to
show~\cite{sander_phase_2013} that $\lrangle{n^l}\sim\sqrt{\Omega^l}$, leading
to $\chi\sim\sqrt{\Omega}$ and $\tilde{\chi}\sim \Omega/N\sim g^{-1}$. Using the
scaling properties of $g$~\cite{mariancutofss},
\begin{equation}
 g\sim\left\lbrace \begin{array}{lll}
       k_c^{3-\gamma} = N^{(3-\gamma)/\omega} & ~~& 2<\gamma<3 \\
       \mbox{const.} & ~~ &\gamma>3
       \end{array}\right.,
\label{eq:g}
\end{equation}
for cutoff scaling as $k_c\sim N^{1/\omega}$, one concludes that, at $\lambda=\lambda_c$, $\chi\sim
N^{\vartheta}$ and $\tilde{\chi}\sim N^{\vartheta'}$ where
$\vartheta=\min[(\gamma-3+\omega)/2\omega,1/2]>0$ and $\vartheta'=\min[(\gamma-3)/\omega,0]\le 0$.
So, the susceptibility $\chi$ always diverges at the transition point while
$\tilde{\chi}$ does not.

\begin{figure}[ht]
 \centering
\includegraphics[width=7cm]{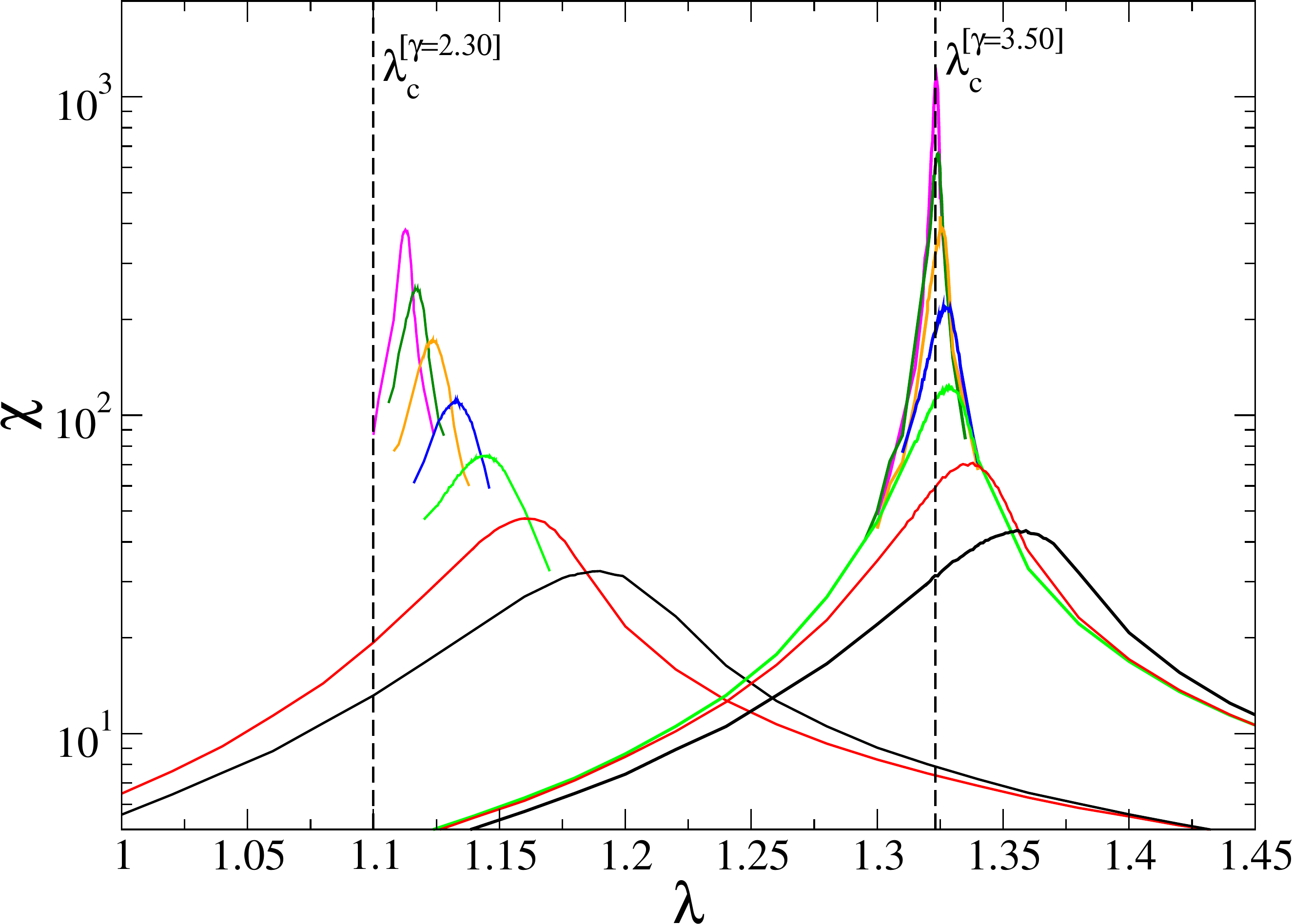}
\includegraphics[width=7.2cm]{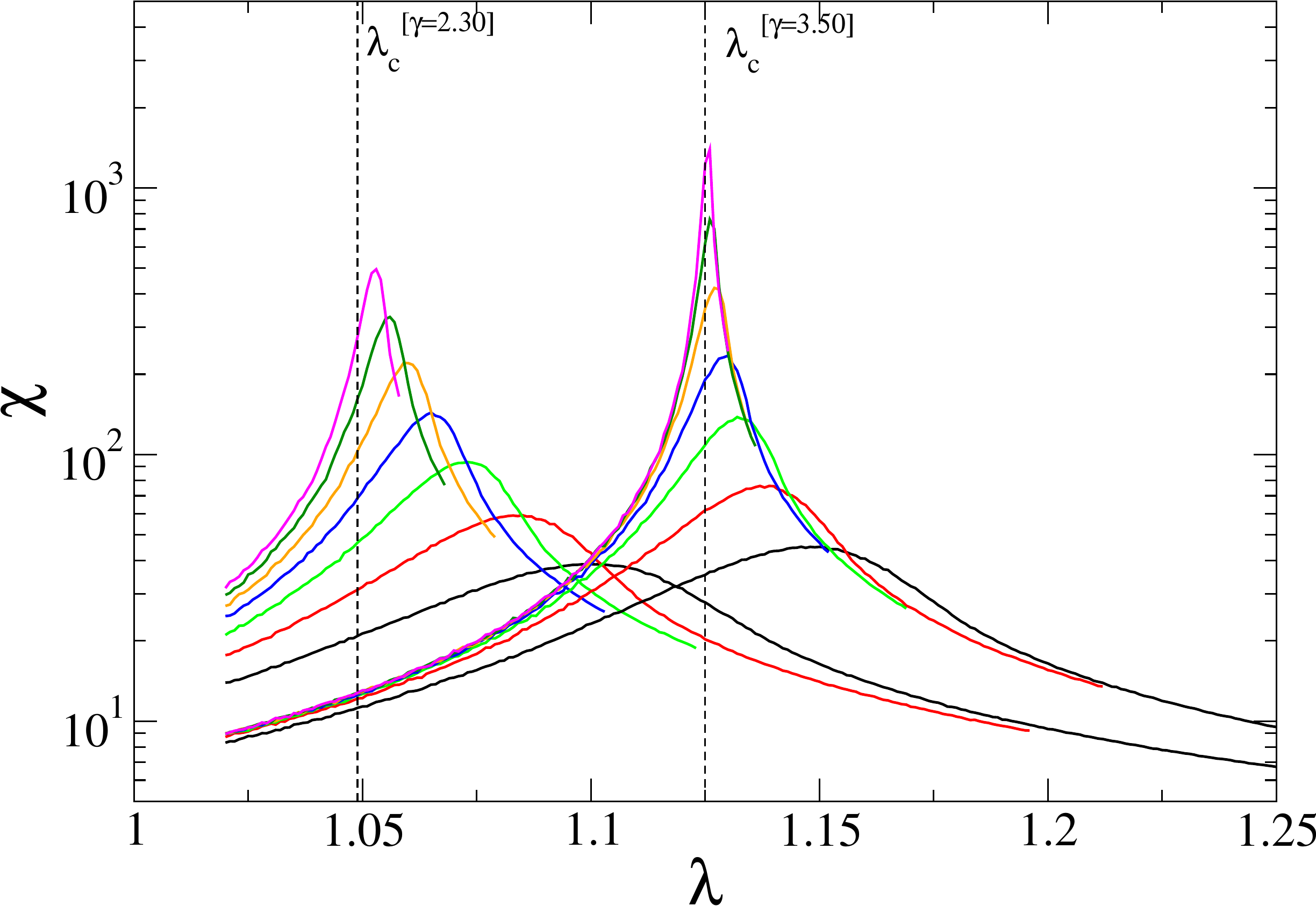}
  \caption{Susceptibility against creation rate for $\gamma=2.30$ (leftmost
curves) and $\gamma=3.50$ (rightmost curves), $k_0=3$ (left) and $k_0=6$ (right). 
The network sizes are $N=10^4,3\times
10^4, 10^5, 3\times 10^5,10^6, 3\times 10^7,10^7$, increasing from  the right. 
Dashed lines are the extrapolations of the peak positions for $N\rightarrow
\infty$}.
 \label{fig:sus}
\end{figure}

Typical susceptibility versus $\lambda$ curves are shown in
Fig.~\ref{fig:sus}. The peak positions shift leftwards converging to a finite
threshold as network size increases. Notice that the larger the degree exponent
the narrower the susceptibility curves and the faster the convergence to  the
asymptotic threshold. The infinite-size threshold $\lambda_c^*$ is estimated in
QS simulations as well as in the mean-field theories using an extrapolation 
\begin{equation}
\lambda_c(N)=\lambda_c^*+a_1N^{-b_1}(1+a_2N^{-b_2}).
\label{eq:nonlin}
\end{equation}
As one can see in Fig.~\ref{fig:lbc}, the curves $\lambda_c$ \textit{vs}. $N$ for
different mean-field theories are only shifted indicating that the exponents
$b_i$ are the same. They can be obtained using a continuous approximation
\begin{equation}
 \lrangle{k}=\int_{k_0}^{k_c}kP(k)dk\simeq\frac{\gamma-1}{\gamma-2}k_0
 \left[1-(k_c/k_0)^{2-\gamma}\right]
\end{equation}
in Eq.~(\ref{eq:lbc_hom}) to obtain $b_1=b_2=(\gamma-2)/\omega$ for $k_c\sim
N^{1/\omega}$, where $\omega=\max(2,\gamma-1)$ for UCM networks~\cite{Catanzaro05}.
These $b_i$ exponents can also be derived directly from equation~(\ref{eq:lbc})
in a more complex calculation that is omitted for sake of brevity.

\begin{table}[ht]
\begin{center}
\begin{tabular}{ccccccccc}\hline
$\gamma$ &&\multicolumn{3}{c}{$k_0=3$}&& \multicolumn{3}{c}{$k_0=6$} \\ \cline{3-5} \cline{7-9}
         &&  PHMF     && $\lambda_c^*$ && PHMF     && $\lambda_c^*$ \\\hline
2.30     && 1.098(1)  && 1.1009(5)   && 1.043(1) && 1.044(1) \\
2.50     && 1.1415(4) && 1.1473(6)   && 1.0628(8)&& 1.0641(5) \\
2.70     && 1.1817(3) && 1.1906(4)   && 1.0788(4)&& 1.0810(7) \\
3.00     && 1.2320(3) && 1.2479(3)   && 1.0977(2)&& 1.1011(4) \\
3.50     && 1.2938(1) && 1.3224(2)   && 1.1200(1)&& 1.1248(4) \\ \hline
\end{tabular}
\end{center}
 \caption{\label{tab:lbc} Transition points of the CP on UCM
networks with different degree exponents, minimum vertex degree $k_0=3$ or
$k_0=6$, and structural cutoff $k_c=N^{1/2}$ for pair heterogeneous mean-field
(PHMF) theory and QS simulations ($\lambda_c^*$). Number in parenthesis are
uncertainties in the last digit.
}
\end{table}

The exponents $b_i$ in QS simulations differ from those of the mean-field 
theories. They can be analytically estimated using the scaling theory presented
in Refs.~\cite{Boguna09,cpannealed}. The QS density at the transition point 
scales as
\begin{equation}
 \bar{\rho}(\lambda_c)\sim (gN)^{-1/2}
\end{equation}
while above it 
\begin{equation}
 \bar{\rho} \sim (\lambda-\lambda_c)^{\beta},
\end{equation}
where $\beta=\max[1,1/(\gamma-2)]$~\cite{Boguna09}. These scaling
laws are confirmed in the pair HMF theory developed in section~\ref{sec:critical}.
Assuming that 
both scaling laws hold at $\lambda_p$ one obtains
\begin{equation}
 \lambda_p-\lambda_c\sim (gN)^{-1/2\beta}.
 \label{eq:lambda_p}
 \end{equation}
Using again the continuous approximation to compute $g$ and neglecting 
higher order terms one finds
\begin{equation}
 g = C_\gamma \times \left\lbrace
 \begin{array}{lll}
 \xi^{3-\gamma}\left[1+\xi^{2-\gamma} +\cdots \right] & ~~ & \gamma<3  \\
 1-\xi^{3-\gamma} +\cdots & ~~ & \gamma>3
 \end{array}   \right.
\end{equation}
where, $\xi\equiv {k_c}/{k_0}$, 
$C_\gamma=|{(\gamma-2)^2}/{(3-\gamma)(\gamma-1)}|$ and a logarithmic 
dependence is found for $\gamma=3$.
Upon substitution of $g$ in Eq.~(\ref{eq:lambda_p}), the exponents 
$b_1=(\gamma-2)(3+\omega-\gamma)/2\omega$ and $b_2=(\gamma-2)/\omega$
for $\gamma<3$ while $b_1=1/2$ and $b_2=(\gamma-3)/\omega$ for $\gamma>3$
and $k_c\sim N^{1/\omega}$are found.

We performed non-linear regressions using Eq.~(\ref{eq:nonlin}) with
$\lambda_c^*$, $a_1$ and $a_2$ free and fixing $b_i$ according to the
theoretical corrections. Excellent fits were obtained, as can be seen in
Fig.~\ref{fig:lbc} and the numerical estimates of $\lambda_c^*$ are shown in
table~\ref{tab:lbc}. As expected, pair HMF theory is a very good
improvement  when compared with the one-vertex approximation $\lambda_c=1$.
However, for some values of $\gamma$ the heuristic HPA theory is
closer to simulations than the pair HMF theory, as can seen in Fig.~\ref{fig:lbc}. 
It is a surprising
result since heterogeneity is expected to play an important role in dynamical
correlations even for degree distributions without a heavy tail as in
the case $\gamma>3$.

The puzzle behind this apparent paradox is that cluster approximations
underestimate the real threshold and the convergence is expected only
in the limit of large cluster approximations. A homogeneous triplet
approximation (HTA) for the CP on unclustered networks yields the 
threshold~\cite{ronan}:
\begin{equation}
\lambda_c =
\frac{\lrangle{k}+2\sqrt{\lrangle{k}^2-\lrangle{k}}}{3\lrangle{k}-4}.
 \label{eq:HTA}
\end{equation}
Comparing this approximation with simulations, figure~\ref{fig:lbc}, one sees
that HTA thresholds are, as expected, higher than the HPA ones but overestimate
the simulation thresholds for all investigated networks, more
evidently for $k_0=3$. This result shows that the homogeneous cluster
approximations will converge to a threshold above the correct one and they are, in principle,
not applicable to the CP dynamics on heterogeneous networks as previously 
done~\cite{JuhaszCP,cpquenched}. The proximity
between HPA theory and simulations is therefore a coincidence.

\section{Critical exponents}
\label{sec:critical}

In this section, the critical exponents of the CP in the pair HMF theory are derived
and compared with results of QS simulations.

\subsection{Critical exponents in the pair HMF theory for infinite networks}

It is well known that cluster approximations of higher orders  improve the
critical point estimates but do change the critical exponents in lattice
systems~\cite{henkel2008non}. As expected, the pair HMF theory for the CP yields
the same scaling exponents as the one-vertex
approximation~\cite{Castellano:2008,cpannealed,Boguna09}, changing only the
amplitudes and the finite-size corrections to the scaling as we will
show in this section. 

In a pair level, the scaling exponents associated to the absorbing state phase
transition can be derived from Eqs.~(\ref{eq:rhok1}) and (\ref{eq:phikk2})
keeping terms up to second order. Assuming again uncorrelated networks, the
dynamical equations become
\begin{equation}
\frac{d\rho_k}{dt} = -\rho_k+\frac{\lambda k}{\lrangle{k}}\sum_{k'}\fkk P(k')
 \label{eq:rhok1_unc}
\end{equation}
and
\begin{eqnarray}
\frac{d\fkk}{dt} & = & -\fkk-\lambda\frac{\fkk}{k'}+\pkk +
\frac{\lambda(k'-1)}{\lrangle{k}}(1+\rho_{k'}
-\rho_k-\fkk)\sum_{k''}\phi_{k'k''}P(k'') \nonumber \\
& & - 
\frac{\lambda(k-1)}{\lrangle{k}}\fkk\sum_{k''}\phi_{kk''}P(k'')+\mathcal{O}(3).
 \label{eq:phikk2_unc}
\end{eqnarray}
The quasi-static approximation with $d\rho_k/dt\approx 0$ and
$d\phi_{kk'}/dt\approx 0$ leads to
\begin{eqnarray}
\fkk  =   \frac{2k'-1}{2k'+\lambda}\rho_{k'}\left\lbrace 1+
\frac{(\lambda+1)(k'-1)}{(2k'-1)(2k'+\lambda)}\rho_k' 
-\right. 
  \left. \left[\frac{k'-1}{2k'-1} + \frac{k'(k-1)}{k(2k'+\lambda)}\right]\rho_k 
\right\rbrace+\mathcal{O}(3)
, \nonumber \\ \label{eq:fkk_2nd}
\end{eqnarray}
which is inserted in Eq.~(\ref{eq:rhok1_unc}) to result
\begin{equation}
 \frac{d\rho_k}{dt} = -\rho_k+\frac{\lambda k}{\lrangle{k}}\left[ \Theta_1
-\rho_k\left(\Theta_2-\frac{\Theta_3}{k}\right)\right]
\label{eq:drhok_s}
\end{equation}
and, consequently, the stationary density
\begin{equation}
 \rho_k = \frac{\lambda k \Theta_1 /\lrangle{k}}{1+\lambda k
\Theta_2/\lrangle{k}-\lambda \Theta_3/\lrangle{k}},
\label{eq:rhok_s}
\end{equation}
where $\Theta_i$ are given by
\begin{equation}
 \Theta_1=\rho - (\lambda+1)\sum_{k}\left[\frac{P(k)\rho_{k}}{(2k+\lambda)} - 
\frac{P(k)(k-1)\rho_{k}^2}{(2k+\lambda)^2} \right] = 
\frac{\rho}{\mathcal{A}_1(\lambda)}+a_1(\lambda)\rho^2,
\label{eq:Theta1}
\end{equation}
\begin{equation}
 \Theta_2=\rho - (\lambda+1)\sum_{k}
\frac{P(k)(3k+\lambda)\rho_{k}}{(2k+\lambda)^2} = 
 \frac{\mathcal{A}_2(\lambda)}{\mathcal{A}_1(\lambda)}
\rho+a_2(\lambda)\rho^2+\cdots
\label{eq:Theta2}
\end{equation}
and
\begin{equation}
 \Theta_3=\sum_{k}\frac{P(k)(2k-1)k\rho_{k}}{(2k+\lambda)^2} = 
\frac{\mathcal{A}_3(\lambda)}{\mathcal{A}_1(\lambda)}\rho+a_3(\lambda)\rho^2,
\label{eq:Theta3}
\end{equation}
where $\rho =\sum_kP(k)\rho_k$ while $A_i$ are constants of order 1 given by
\begin{equation}
 \mathcal{A}_1(\lambda) =
1+\frac{\lambda(\lambda+1)}{\lrangle{k}}\sum_k\frac{kP(k)}{2k+\lambda},
\end{equation}
\begin{equation}
 \mathcal{A}_2(\lambda) =
1-\frac{\lambda(\lambda+1)}{\lrangle{k}}\sum_k\frac{k^2P(k)}{(2k+\lambda)^2}
\end{equation}
and
\begin{equation}
 \mathcal{A}_3(\lambda)=\frac{\lambda}{\lrangle{k}}\sum_k\frac{k^2(2k-1)P(k)}{
(2k+\lambda)^2}.
\end{equation}
The rightmost sides of Eqs.~(\ref{eq:Theta1})-(\ref{eq:Theta3})
were obtained  using $\Theta_i<\rho$  (the proofs of these bounds are simple) 
and Eq.~(\ref{eq:rhok_s}) in a self-consistent iterative approach~\cite{barratbook}. 
The constants $a_i$ are of order $1/\lrangle{k}^2$ and their explicit
forms are omitted.

Multiplying Eq.~(\ref{eq:drhok_s}) by $P(k)$ and summing over $k$ 
$(k_c\rightarrow\infty)$ one finds
\begin{equation}
 \frac{d\rho}{dt} =
-\rho+\lambda\Theta_1+\frac{\lambda\Theta_3\rho}{\lrangle{k}} - 
\frac{\lambda\Theta_2}{\lrangle{k}}\lrangle{k\rho_k}
\label{eq:drhoself0}
\end{equation}
where
\begin{eqnarray}
 \lrangle{k\rho_k}& = &(\gamma-1)k_0^{\gamma-1}\psi_1\int_{k_0}^\infty 
\frac{k^{-\gamma+2}}{1+\psi_2k}dk \nonumber \\
& = & 
\frac{\gamma-1}{\gamma-2}\frac{\psi_1}{\psi_2}
F\left(1,\gamma-2,\gamma-1,\frac{-1}{\psi_2k_0}\right),
\end{eqnarray}
$\psi_i=\lambda\Theta_i/[\lrangle{k}-\lambda\Theta_3]$ and $F(a,b,c,x)$ is 
the Gauss hypergeometric function~\cite{gradshteyn2007}. Near to the critical 
point, $\psi_i\ll 1$, we can use the asymptotic form of $F(a,b,c,x)$ to finally
find
\begin{equation}
 \frac{d\rho}{dt}= -\rho+\frac{\lambda}{\mathcal{A}_1}\rho -
\tilde{\alpha}_1\rho^2
- \tilde{\alpha}_2\rho^{\gamma-1}+\cdots,
\label{eq:drhoself}
\end{equation}
where $\tilde{\alpha}_i(\lambda)$, $i=1,2$, are positive parameters whose
details are 
omitted for sake of conciseness.

The stationary
density close to the transition point is given by
\begin{equation}
\tilde{\alpha}_1\rho+\tilde{\alpha}_2\rho^{\gamma-2} \simeq
\frac{\lambda-\mathcal{A}_1}{\mathcal{A}_1}.
\end{equation}
An expansion around $\lambda=\lambda_c$ yields 
\begin{equation}
 \mathcal{A}_1(\lambda)=\lambda_c+(\lambda-\lambda_c)\mathcal{A}'_1(\lambda_c)+\cdots
\end{equation}
where the identity $\lambda_c=\mathcal{A}_1(\lambda_c)$ comes from
Eq.~(\ref{eq:lbc}).
Considering only the leading term in $\rho$ one finds
\begin{equation}
\bar{\rho}\sim
(\lambda-\lambda_c)^{\beta},~~~~~\beta=\max\left[1,\frac{1}{\gamma-2}\right].
\end{equation}

At the transition point $\lambda=\lambda_c$, equation~(\ref{eq:drhoself}) 
becomes
\begin{equation}
 \frac{d\rho}{dt}= - \tilde{\alpha}_1\rho^2
- \tilde{\alpha}_2\rho^{\gamma-1},
\end{equation}
which yields $\rho\sim t^{-\delta}$ where $\delta=\beta=\max[1,1/(\gamma-2)]$.
Finally, close to the critical point one can show that 
\begin{equation}
 \rho-\bar{\rho}\sim
\exp\left[-\left(\frac{\lambda-\mathcal{A}_1}{\mathcal{A}_1}\right)t\right]
\end{equation}
leading to a relaxation time scaling as 
\begin{equation}
 \tau=\frac{\mathcal{A}_1}{\lambda-\mathcal{A}_1}\sim
(\lambda-\lambda_c)^{-\nu_\parallel}
\end{equation}
with  a $\gamma$-independent exponent $\nu_\parallel=1$. The exponents $(\beta,\delta,\nu_\parallel)$
obtained in this section are exactly the same of the one-vertex HMF theory~\cite{Castellano:2006}.

\subsection{Finite-size scaling critical exponents}

 Finite-size scaling (FSS) exponents associated to the QS state can be obtained
using a mapping of the CP dynamics in a one-step process~\cite{vankampen} as
proposed in Ref.~\cite{Castellano:2008}. For finite-size systems the condition
$\rho k_c\ll 1$ is applicable for long times  and very close to the transition
point. So, we approximate Eq.~(\ref{eq:rhok_s}) by 
$\rho_k\simeq \lambda k \Theta_1/\lrangle{k}$ which is inserted in
Eq.~(\ref{eq:drhoself0}) to find
\begin{equation}
 \frac{d\rho}{dt} = -\rho+\frac{\lambda}{\mathcal{A}_1}(1-\tilde{g}\rho)\rho
 \label{eq:drhodt_gtil}
\end{equation}
where the factor $\tilde{g}$ is given by
\begin{equation}
 \tilde{g} =
\frac{\lambda\mathcal{A}_2}{\mathcal{A}_1}\frac{\lrangle{k^2}}{\lrangle{k}^2}
-\mathcal{A}_1a_1-\frac{\mathcal{A}_3}{\lrangle{k}}.
 \label{eq:gtil}
\end{equation}

The first term proportional to $\rho$ in Eq.~(\ref{eq:drhodt_gtil}) represents 
an annihilation $n\rightarrow n-1$ whereas
the second one a creation event $n\rightarrow n+1$. 
Following the interpretation of Ref.~\cite{Castellano:2008},  in a mean-field
level Eq.~(\ref{eq:drhodt_gtil}) represents a one-step process defined by a
transition rate $W(n,m)$ from a state with $m$ to another with $n$ particles
given by
\begin{equation}
  W(m,n) = n\delta_{m,n-1} + \frac{\lambda}{\mathcal{A}_1}(1-\tilde{g}\rho)n
\delta_{m,n+1}.
\end{equation}
At the critical point, we have the additional simplification
$\lambda_c=\mathcal{A}_1$ and the 
transition rate becomes equal to that of the one-step process
associated to the CP dynamics in a one-vertex 
HMF theory~\cite{Castellano:2008}, with the factor
$g=\lrangle{k^2}/\lrangle{k}^2$ 
replaced by $\tilde{g}$, given by Eq.~(\ref{eq:gtil}). The QS analysis  of this 
critical one-step process with the original $g$ factor was 
done in Ref.~\cite{cpannealed}, whose results are 
presented below. 

The  QS probability distribution $P(n)$ is given by Eq.~(\ref{eq:pofn})
with $\Omega = N/\tilde{g}$. The QS density $\bar{\rho}$ and the characteristic 
time $\tau$, defined as
$\bar{\rho}=\frac{1}{N}\sum_nnP(n)$ and $\tau=1/P(1)$~\cite{Mancebo2005},
respectively, 
scale as
\begin{equation}
 \rho\sim (\tilde{g}N)^{-1/2} ~~~~\mbox{and}~~~~~ \tau\sim (N/\tilde{g})^{1/2}.
\label{eq:rhotauqs}
\end{equation}
Nevertheless,  the factor $\tilde{g}$ has exactly the same asymptotic scaling
properties as the factor $g$, which are given by Eq.~(\ref{eq:g}), and therefore
the  same FSS exponents of the one-vertex HMF are obtained in  pair HMF
approximation. The scaling laws $\bar{\rho}\sim N^{-\nu}$ and $\tau\sim
N^{\alpha}$ with $\nu=\max[(5-\gamma)/2,1/2]$ and
$\alpha=\max[(\gamma-1)/4,1/2]$ are obtained for UCM networks with a structural cutoff
$k_c\sim N^{1/2}$~\cite{Catanzaro05}.

\begin{table}[ht]
\begin{center}
\begin{tabular}{ccccccccc}
\hline
$\gamma$ &&\multicolumn{3}{c}{$k_0=3$}&& \multicolumn{3}{c}{$k_0=6$} \\ \cline{3-5} \cline{7-9}
         &&   $S_\nu$ && $S_\alpha$  &&  $S_\nu$ && $S_\alpha$ \\\hline
2.3  && 0.50(2) && 0.48(2)&& 0.50(1) && 0.50(1)\\
2.5  && 0.51(2) && 0.47(2)&& 0.50(1) && 0.51(1)\\
2.7  && 0.51(2) && 0.49(2)&& 0.50(1) && 0.50(1)\\
3.0  && 0.51(2) && 0.49(2)&& 0.50(1) && 0.50(1)\\
3.5  && 0.51(2) && 0.48(2)&& 0.51(1) && 0.50(1)\\ \hline
\end{tabular}
\end{center}
 \caption{\label{tab:exp}Critical exponents obtained in QS simulations of the CP
on UCM networks with minimum degrees $k_0=3$ or $k_0=6$ and cutoff $k_c=N^{1/2}$.
The exponents were obtained in power law regressions
$\bar{\rho}\sim(\tilde{g}N)^{-S_\nu}$ and $\tau\sim (N/\tilde{g})^{S_\alpha}$.}
\end{table}

\begin{figure}[ht]
 \centering
\includegraphics[width=8cm]{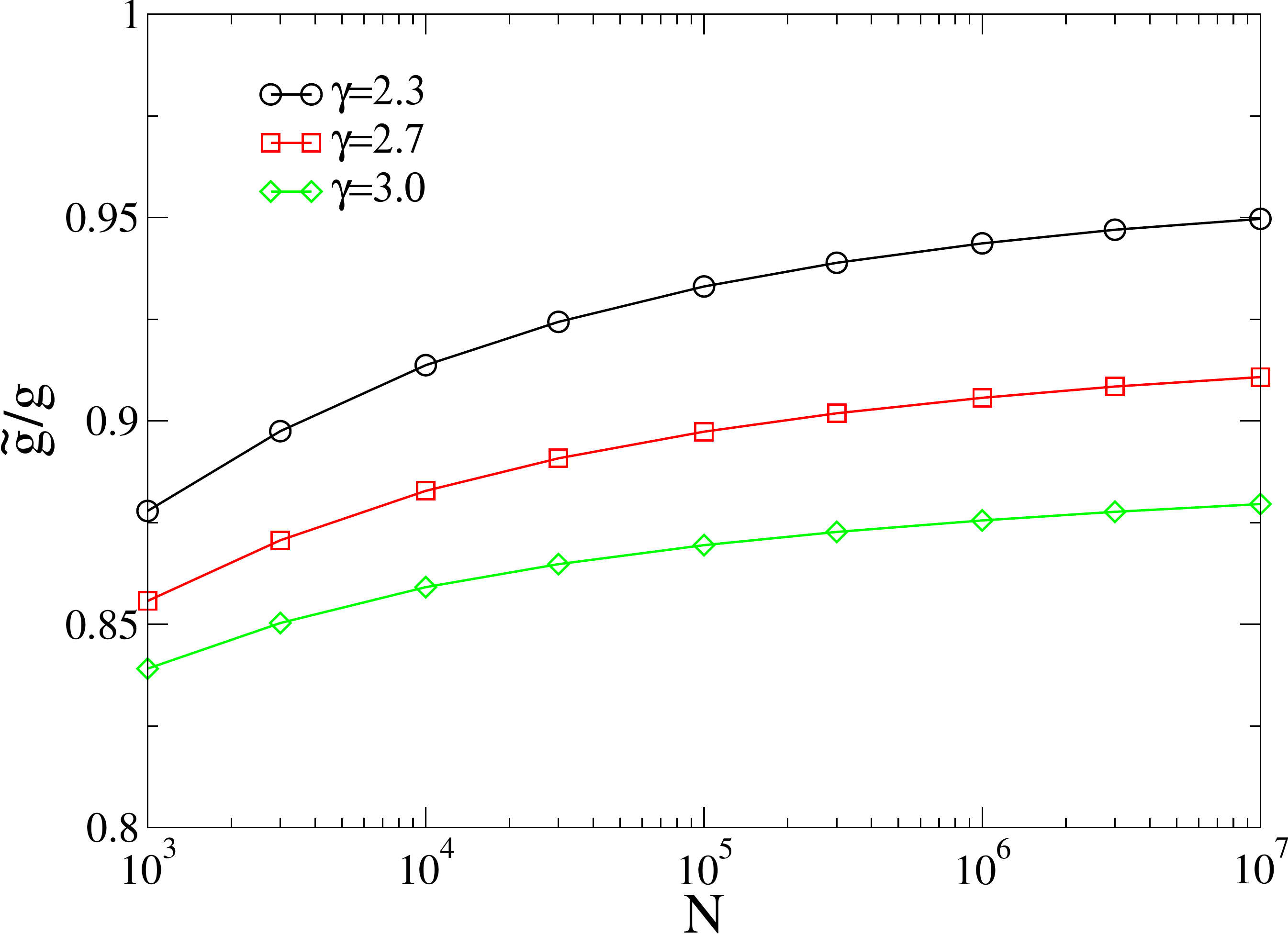}                                  
            
 \caption{Ratio between the factor $\tilde{g}$ obtained in
pair HMF theory, Eq.~(\ref{eq:gtil})  with $\lambda=\lambda_c(N)$, 
and the factor $g=\lrangle{k^2}/\lrangle{k}^2$ of the one-vertex HMF theory, for $k_0=3$.}
 \label{fig:gtil}
\end{figure}

\begin{figure}[ht]
 \centering
~~  \includegraphics[width=7cm]{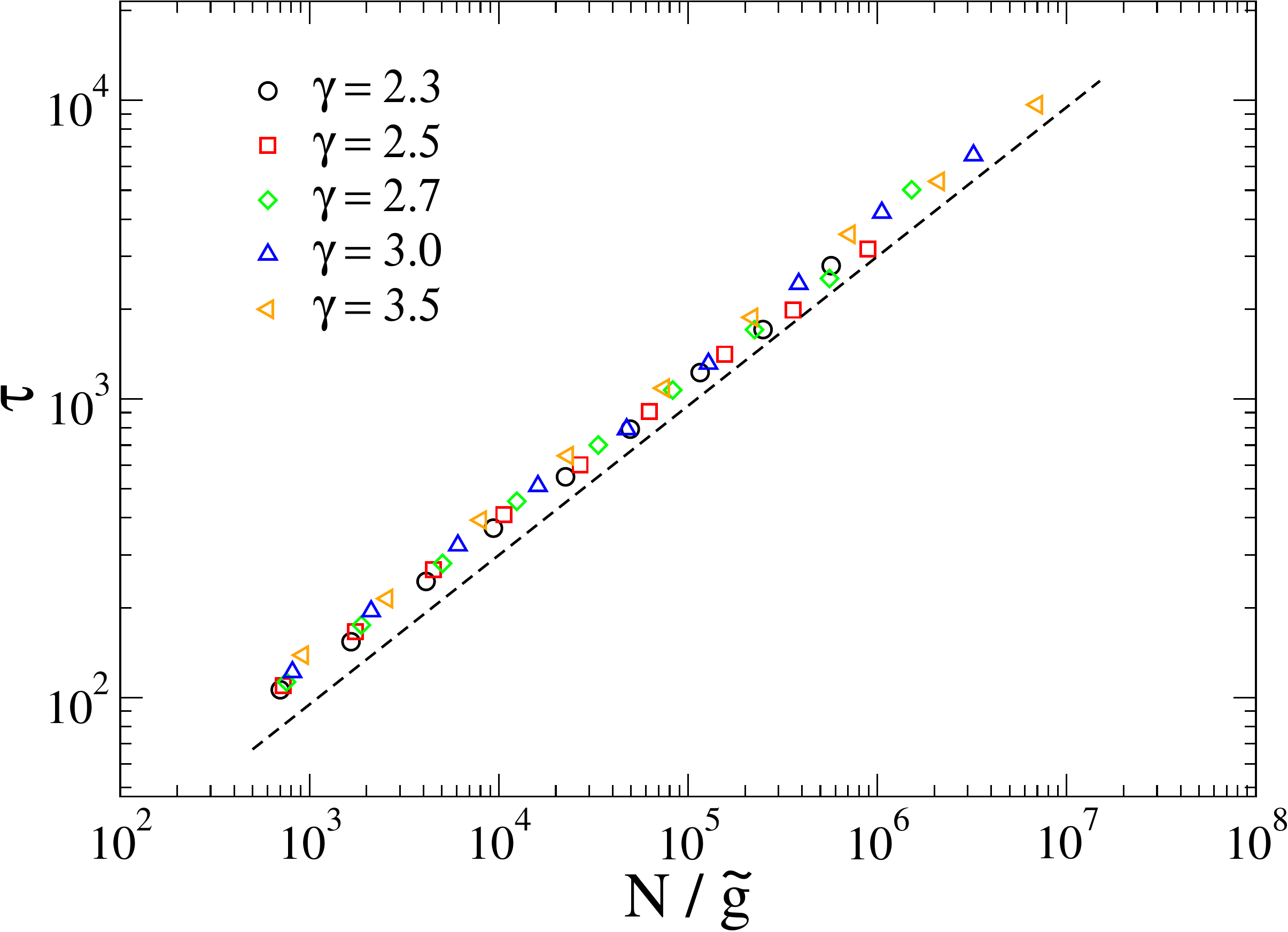}     
~~  \includegraphics[width=7cm]{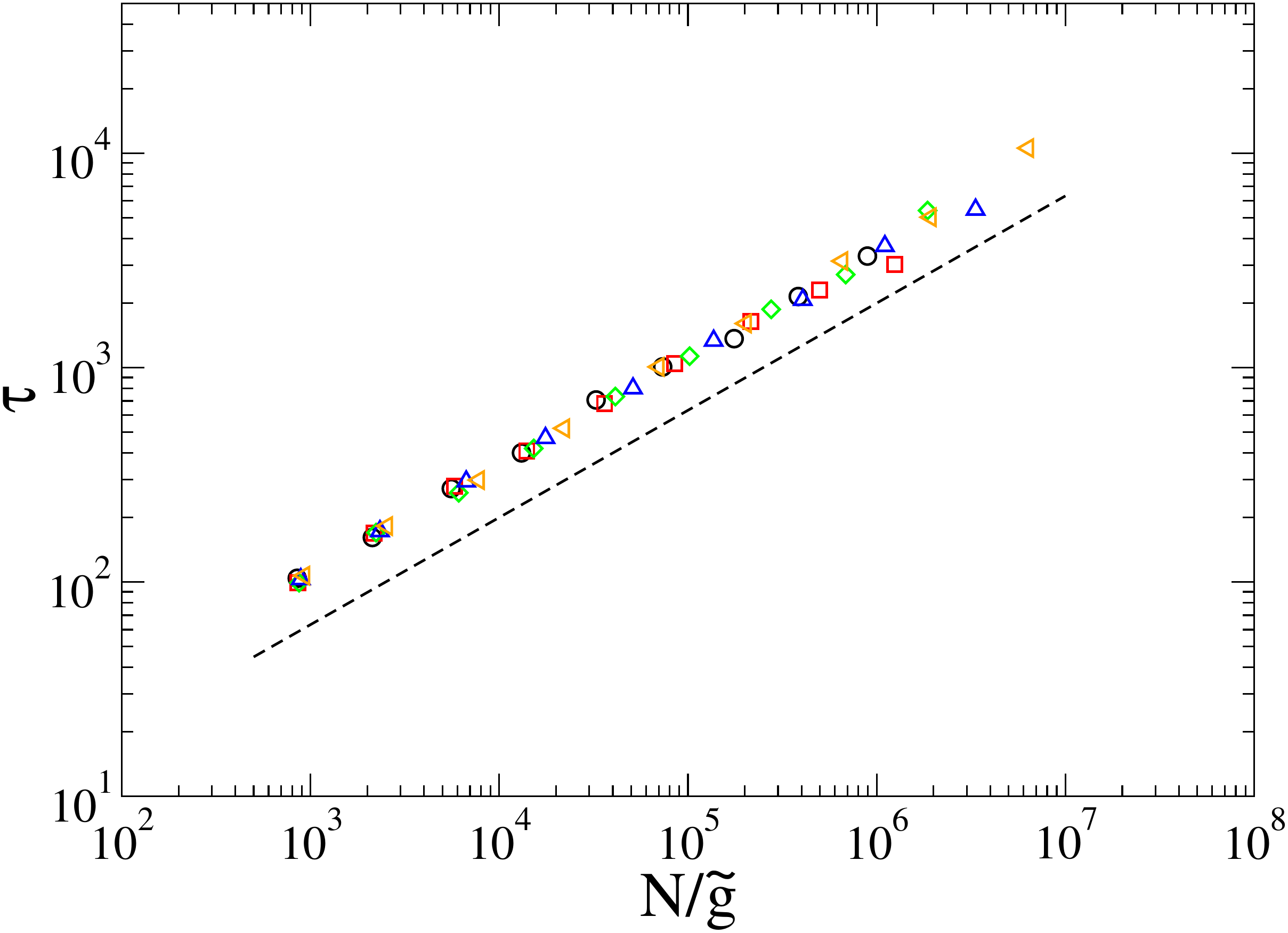} 

~\\

~~\includegraphics[width=7cm]{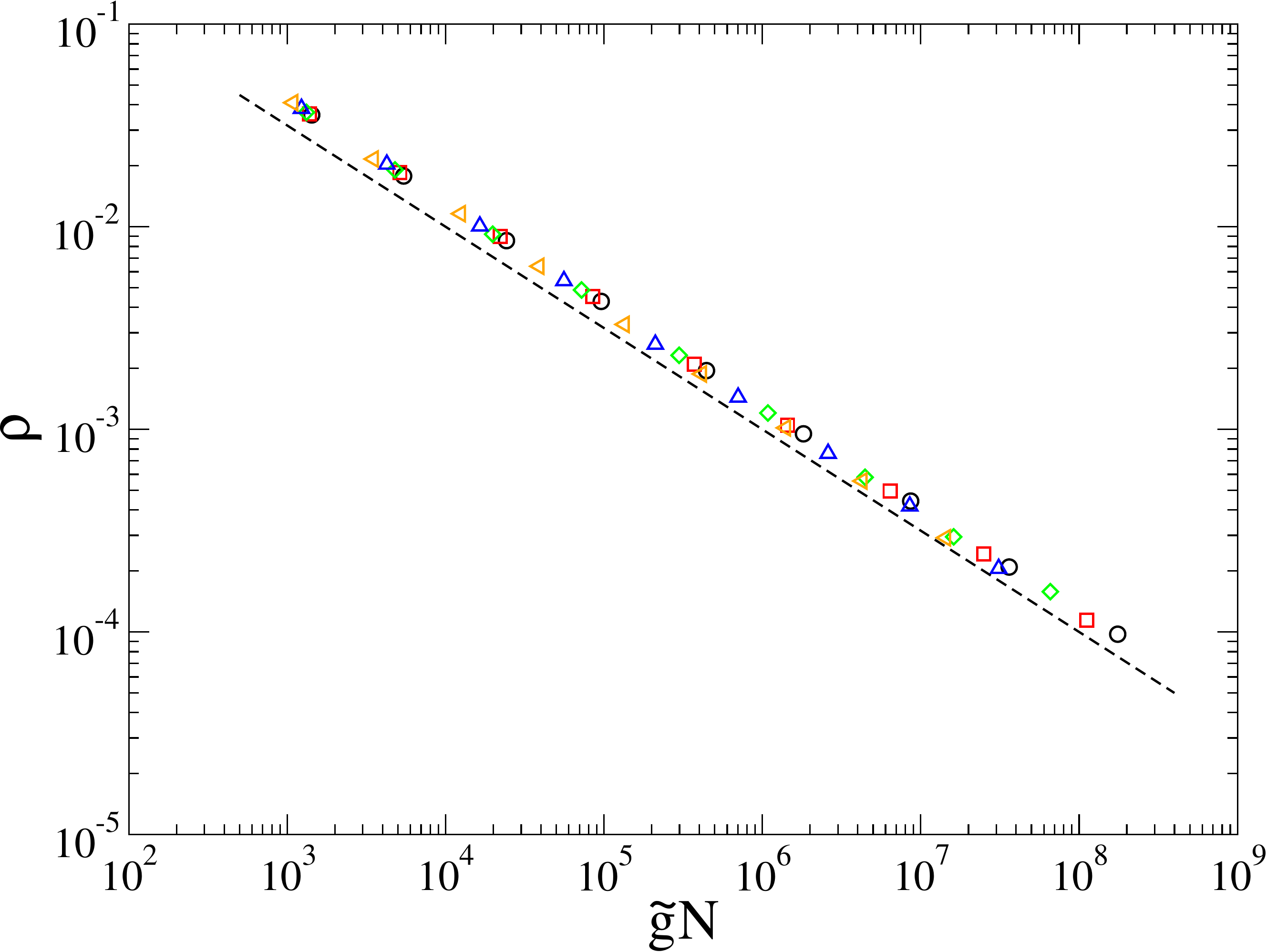}
~~ \includegraphics[width=7cm]{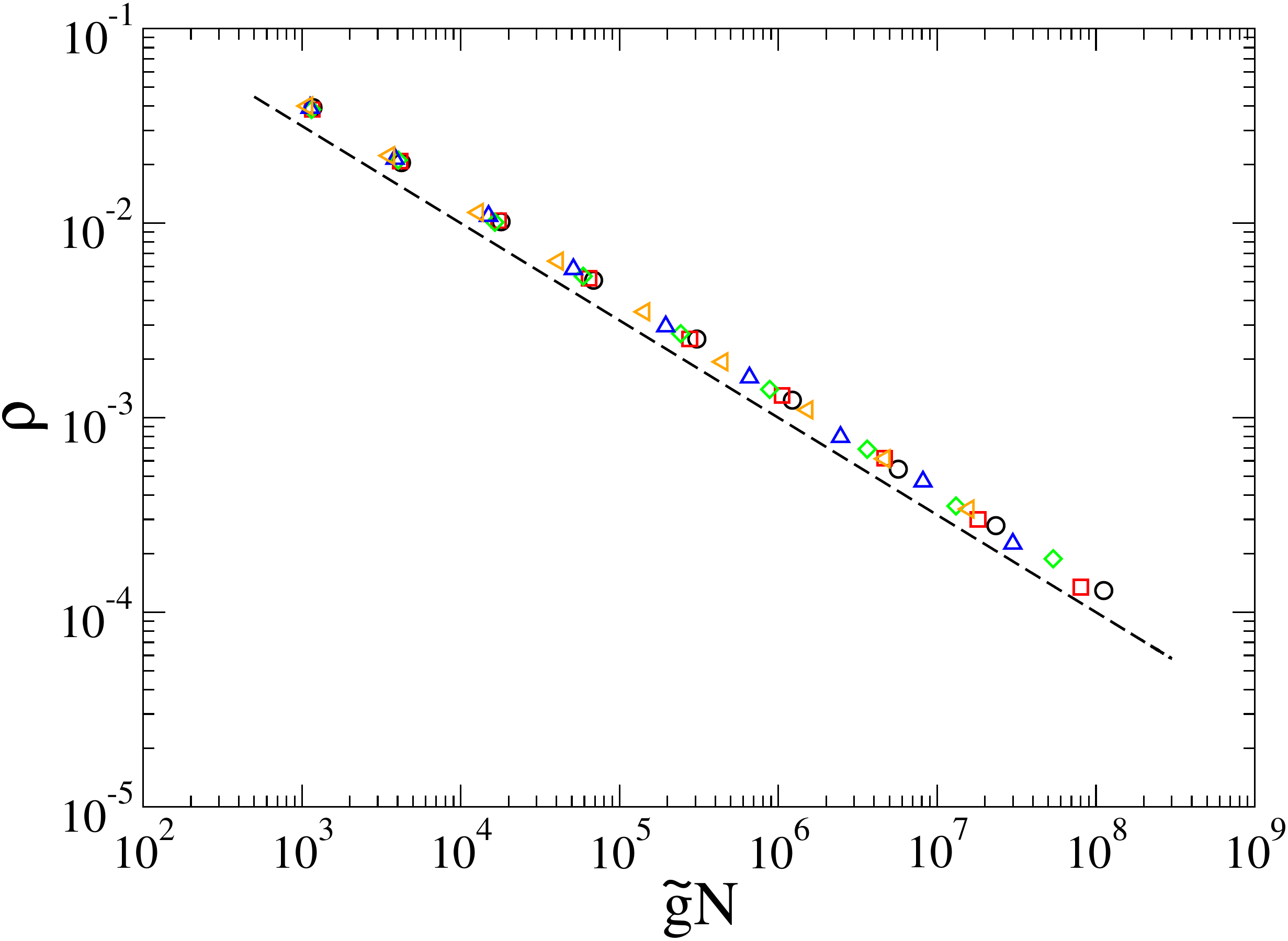} 
\caption{\label{fig:qscrit}FSS of the characteristic time  and critical QS
density for $k_0=3$ (left) and $k_0=6$ (right). The dashed lines have slope 
$\pm$1/2 as guides to the
eyes.}
\end{figure}

Despite of the same asymptotic scaling, the sub-leading corrections in the new
factor $\tilde{g}$ are not negligible as one can see in
Fig.~\ref{fig:gtil}.
Moreover, the finite-size corrections in the critical point position observed
for pair HMF theory as well as in QS simulations (Fig.~\ref{fig:lbc})
suggest that we must compute the critical quantities at $\lambda_p(N)$ and not
$\lambda_c^*$ as previously done~\cite{cpquenched}.
Figure~\ref{fig:qscrit} shows double-logarithmic plots for the FSS of the critical QS
density and characteristic time following this strategy. For the wide range of 
degree exponents analyzed, the values obtained from power law regressions
$\bar{\rho}\sim(\tilde{g}N)^{-S_\nu}$ and $\tau\sim (N/\tilde{g})^{S_\alpha}$
are in remarkable agreement with the theoretical prediction $S_\nu=S_\alpha=1/2$, as one can
verify in table~\ref{tab:exp}. Most importantly, the scaling laws hold for the
entire range of investigated sizes in contrast with the analysis for a fixed
$\lambda=\lambda_p^*$ and using the old factor $g$, for which large deviations of
the theoretical scaling laws are observed at
small sizes, the more evident for more heterogeneous networks ($\gamma\le
2.5$)~\cite{cpquenched}. Noticeably, the exponent of the characteristic time
for $\gamma=2.3$ is in great agreement with the theory if factor $\tilde{g}$ is
used in contrast with a poor accordance observed for a similar degree exponent
reported in Ref.~\cite{cpquenched}. It is worth stressing that the almost
perfect match is found only if both factor $\tilde{g}$ and corrections in
$\lambda_p(N)$  are used concomitantly. In particular, for the $k_0=3$ case the 
scaling laws obtained in simulations are not consistent with HMF if this strategy
is not used.  Thus, we filled a missing gap showing
that the critical exponents as well as the sub-leading corrections to the FSS are
very accurately predicted by the pair HMF theory.

\section{Conclusions}
\label{sec:conclu}

The dynamics of the contact process on the top of complex networks was
investigated using a pair heterogeneous  mean-field theory in which the vertices
are grouped accordingly their degrees. We compared the theoretical results with
QS simulations and showed that they represent great improvements in relation to the simple
HMF approach. However, for a wide range of the degree distributions, a heuristic
homogeneous pair-approximation~\cite{JuhaszCP,cpquenched} is still more
accurate than our heterogeneous approach. To unveil this contradiction we
compared simulations with a homogeneous triplet  approximation 
that must be more accurate than homogeneous pair-approximations. We observed, however, that
the HTA theory overestimates the simulation thresholds showing that successive
homogeneous cluster approximations~\cite{Avraham92} converge to the wrong critical point
and, therefore, that the agreement between HPA and simulations is only a
coincidence.

We also determined the critical exponents in the pair HMF approach. For the
infinite size limit the exponents are the same as the one-vertex theory. 
However, the finite-size corrections to the scaling obtained in the
pair HMF theory allowed a remarkable agreement with QS simulations for all
degree exponents ($2.3\le \gamma\le3.5$) and network sizes ($10^3\le N\le 10^7$)
investigated, suppressing a deviation observed for low degree exponents in the
one-vertex HMF theory~\cite{cpquenched}. Our results strongly corroborate that
HMF theories predict the correct scaling exponents of the CP on SF random
networks.

The present theoretical approach can be applied to other important
dynamical processes on complex networks as the generalized voter
models~\cite{Moretti}, sandpiles~\cite{Goh} as well as more sophisticated
structures as multiscale and multiplex network~\cite{Mucha,Gomez13}. Our
approach permits to explicitly derive analytical expressions whereas previous 
pair-approximations for dynamical processes in complex
networks~\cite{Gleeson2011,Pugliese09} usually need a numerical integration of the
corresponding master equations, which limits the analysis to relatively
smaller systems. As an example, the  threshold of the SIS model in a pair HMF
approximation can easily obtained:  
\begin{equation}
\lambda_c = \frac{\lrangle{k}}{\lrangle{k^2}-\lrangle{k}}.
\end{equation}
This threshold coincides with that of the susceptible-infected-recovered (SIR)
model in a one-vertex HMF theory~\cite{barratbook}. This results was recently
proposed in Ref.~\cite{boguna2013nature} using heuristic arguments.

The pair HMF theory is different from other pair approximations 
for networked systems~\cite{mata2013pair,Pugliese09,Gleeson2013,Gleeson2011}.
However, the pair HMF can be obtained from pair QMF~\cite{mata2013pair} 
performing a coarse-graining where vertices and pair 
are grouped according to their degrees. We also performed the pair QMF analysis
for CP and found thresholds slightly below pairs HMF, but we could not determine 
the scaling exponents in this approach. As a prospect, it would be interesting to 
perform numerical integration of Eqs.~(\ref{eq:rhok1}) and (\ref{eq:phikk2})
in a nonperturbative analysis for a comparison  with the nonperturbative HMF~\cite{Gomez13}  
and the general pair approximation for binary states~\cite{Gleeson2013,Gleeson2011}.

\section*{Acknowledgments}
This work was partially supported by the Brazilian agencies CNPq and
FAPEMIG. ASM thanks the financial support from CAPES. RSF acknowledges 
financial support under project MULTIPLEX, European Commission, FET 
Proactive IP Project number 317532.

\bigskip

\providecommand{\newblock}{}

\end{document}